\documentclass{article}
\usepackage{spconf,amsmath,graphicx}
\usepackage[utf8]{inputenc}
\usepackage{amsmath}
\usepackage{algorithm}
\usepackage[noend]{algpseudocode}

\title{Multiscale edge detection and parametric shape modeling for boundary delineation in optoacoustic images}
\name{S Mandal$^{\star \dagger}$, Viswanath PS $^{\star}$, Yeshaswini N $^{\star}$, 
 XL De{\'a}n-Ben $^{\dagger}$, D Razansky$^{\star \dagger}$}

\address{$^{\star}$ Technische Universit\"{a}t M\"{u}nchen, Arcisstraße 21, 80333 M\"{u}nchen, Germany \\
         $^{\dagger}$ Helmholtz Zentrum M\"{u}nchen, Ingolstädter Landstr 1, 85764 Neuherberg, Germany\\
         email:\{s.mandal,dr\}@tum.de}

\begin{document}
\ninept
\maketitle
\begin{abstract}
In this article, we present a novel scheme for segmenting the image boundary (with the background) in optoacoustic small animal \textit{in vivo} imaging systems. The method utilizes a multiscale edge detection algorithm to generate a binary edge map. A scale dependent morphological operation is employed to clean spurious edges. Thereafter, an ellipse is fitted to the edge map through constrained parametric transformations and iterative goodness of fit calculations. The method delimits the tissue edges through the curve fitting model, which has shown high levels of accuracy. Thus, this method enables segmentation of optoacoutic images with minimal human intervention, by eliminating need of scale selection for multiscale processing and seed point determination for contour mapping. 
\end{abstract}
\begin{keywords}
Image segmentation, Photoacoustic effects, Image edge detection, Curve fitting 
\end{keywords}
\section{Introduction}
\label{sec:intro}
Optoacoustic Tomography (OAT) has emerged in the last decade as a new hybrid bioimaging modality combining the advantages of acoustical detection and optical absorption contrast. OAT provides structural as well as functional information of biological tissues in two (2D) and three (3D) dimensions, with resolution and frame rates representative of ultrasound \cite{razansky2011volumetric}. Research efforts in OAT have been directed towards the development of new hardware components and inversion methodologies allowing increasing imaging speed and resolution, as well as on investigating potential biomedical applications. Recently developed OAT cross-sectional systems enable whole-body small animal \textit{in-vivo} imaging. The unique capabilities offered open up the unexplored domain of post-reconstruction image analysis for this modality. Segmentation is one the most challenging tasks in image processing, and the relatively low intrinsic contrast of background structures (compared to digital photography) and limited view problems of OAT increase the complexity involved. Classical experiments on non-human primates demonstrate that complex objects are perceived by vision system in a multiscale manner, and are tuned from coarse scale to finer scales \cite{sumengen2005multi}, we use the information as a basic assumption in our study. For the edge detection at individual scales, the Sobel operator, which approximates the gradient of the image intensity function, or the Canny edge detector \cite{canny1986computational} which uses a feature synthesis step from fine to coarse scales, are commonly employed. Multiscale edge detection, a rich area of interest in itself, can enhance the performance of edge detection in a fixed (original) scale. For example, the Perona-Malik flow (anisotropic diffusion)\cite{perona1990scale} gives a successful formulation of scale-space adaptive smoothing function, leading to better preservation of edges while achieving a better noise performance. These edge detection methods have been demonstrated to be useful for optoacoustic images and employed for calibration of reconstruction parameters by Mandal et.al.\cite{Mandal2014128}. More sophisticated techniques were proposed by Tabb and Ahuja \cite{tabb1997multiscale} and followed by Ma and Manjunath \cite{ma2000edgeflow}who developed the concept of designing a vector field for edge detection. In \cite{tabb1997multiscale} the vector field is created by analyzing the neighborhood of a pixel, where the size and spatial scale of the neighborhood were determined heuristically by an homogeneity parameter, and is adaptively determined on a pixel to pixel basis. The edge flow method suggested by \cite{ma2000edgeflow} utilizes the color and textural information of images to track changes in directions, creating a vector flow. This method detects boundaries when there are two opposite directions of flow at a given location in a stable state. However, it depends strongly on color information and requires a user defined scale to be input as a control parameter. In a follow up work by \cite{sumengen2005multi}, a methods was demonstrated which eliminates the need of scale selection and potentially works with gray-scale images. We have developed a  new segmentation method for OAT by integrating the multiscale edgeflow, scale space diffusion and morphological image processing. The suggested multiscale edge detection method is illustrated in section 2, and the processing results of OAT images are displayed in section 3. In section 2.4, we discuss about delineating the tissue boundaries in OAT, which is another challenging issue we aim to tackle. Due to illumination conditions and limited view considerations in the detection, the reconstructed images can be affected by open edges and fuzzy boundaries. Traditional edge linking algorithms may then fail to provide satisfactory segmentation. Thereby, we suggest a novel curve fitting model that delimits the tissue-background boundaries in OAT images. We further characterize the goodness of fit (GoF) by using Dice coefficient metric (DM), an standard evaluation matrix. The GoF is employed to iteratively improve the segmentation performance. 
 
\section{Methods and Algorithms}
\label{sec:multiseg}

The proposed methodology aims at providing a framework of OAT image segmentation and reducing the parameters that need to be defined to achieve a segmented boundary between imaged biological tissue and acoustical coupling medium. Details on the imaging setup and the implementation steps of the algorithm are given in the following subsections.

\subsection{Imaging Setup and Protocol}
\label{ssec:imaging}

A commercial whole-body small animal multispectral optoacoustic tomography (MSOT) scanner (Model MSOT256-TF, iTheraMedical GmbH, Munich, Germany) was used for data acquisition. The scanner consists of a 256-element array of cylindrically focused piezocomposite transducers with 5 MHz central frequency. The transducer array covers an angle of approximately 270 degrees and has a radius of curvature of 40 mm. Light excitation is provided with the output laser beam from a wavelength-tunable optical parametric oscillator (OPO)-based laser, which is shaped to attain ring-type uniform illumination on the surface of the animal (mouse) by means of a fiber bundle with an output arranged on annular illumination slots. The system allows simultaneous acquisition of the signals generated with each laser pulse, which are digitized at 40 MS/s. Thereby, the scanner is capable of rendering 10 cross-sectional images per second. For acquisition of the entire mouse cross-sections the laser wavelength was set between 690 and 900 nm and the signals were averaged 10 times in order to improve SNR performance. The optoacoustic images representing the distribution of optical absorption were reconstructed using a non-negative constrained model-based inversion algorithm \cite{6748955}. Prior to reconstruction, the signals were initially band-pass filtered with cut-off frequencies between 0.1 and 7 MHz for removing low-frequency offsets and high-frequency noise. The reconstructed images have a 200 x 200 effective pixel resolution (for a reconstruction area of 20 x 20 mm$^{2}$) and are represented in gray-scale (normalized 0-255 level for image processing applications). The animal handling for imaging applications were performed in full conformity with institutional guidelines and with approval from the Government of Upper Bavaria (Germany).

\subsection{Edge-flow vector field and edge detection}
\label{sse: edgeflow}

The general trend in multiscale edge detection is to define a scale a-priori and then estimate the scale locally, however, Sumengen and Manjunath \cite{sumengen2005multi} suggested a geometrically inspired method that estimates the edges that exists both in coarse and fine scales, and localize them in the fine scale. In this article, we propose a modified version of the edgeflow methods \cite{ma2000edgeflow, sumengen2005multi}, by integrating anisotropic diffusion and scale-space dependent morphological processing for convenient application on OAT images. The edge flow algorithm defines a vector field, such that the vector flow is always directed towards the boundary on both its sides. The classical edge flow model utilizes a vector propagation stage. In the current method, the relative directional differences are considered for computing gradient vector. The gradient vector strengthens the edge locations and tracks the direction of the flow along x and y directions. The search function looks for sharp changes from positive to negative signs of flow directions and whenever it encounters such changes, the pixel is labeled as an edge point. The magnitude of the change is the deciding factor behind the edge strength, which is reflected as edge intensity in the final edge map. The vector field is generated explicitly from fine to coarse scales, whereas the multiscale vector conduction is implicitly from coarse to finer scales. Thus, the algorithm is suitable for localizing the edges in the finer scales, which is achieved by preserving only the edges and neighborhoods that exists in several scales (depending on the threshold employed), and suppressing features that disappear rapidly with increment of scales. The workflows for the vector field generation and morphological processing are illustrated in the Algorithm 1. The pseudocode further outlines the process of geometric curve fitting. In the current implementation, the Gaussian offset ($\sigma$) is also lowered for the gray-scale OAT images. For the acquired in vivo OAT images, optimum performance was achieved with $\sigma = 3$. We analyze the images between scales s=1, and s=3, where s=1 is the starting (finest) scale. The interval is sampled at sub-pixel resolution [$ \Delta s= 0.5$] for tracing the dislocation of edges in subsequent edges, as used by Berghold \cite{10.1109/TPAMI.1987.4767980}.

\makeatletter
\def\BState{\State\hskip-\ALG@thistlm}
\makeatother

\begin{algorithm} [htb]
\caption{Multiscale Edge Detection and Curve Fit}\label{euclid}
\begin{algorithmic}
%\Procedure{MyProcedure}{}
\State $\textit{I(x,y)} \gets \text{image}$
\State $\textit{s}_1,\textit{s}_n \gets \text{smallest and largest scale of the image respectively}$
\State $\Delta s =0.5$ \text{ be the sampling interval for each scale}
\State $s ={s}_1$ \Comment{ initialize book keeping}
\State ${\vec{U}},{\vec{U}}_{new} \gets \nabla I(x,y)$ \Comment{Gradient Vector}
\State ${\vec{U}}_{new} \gets \nabla c.\nabla I(x,y)+c(x,y,t)\Delta I$ \Comment{Anisotropic Diffusion}
\While{${s}_1 < {s}_n }$
\State $s \gets s + \Delta s$
\State $M \gets max(\parallel {\vec{U}} \parallel)$
\State ${\vec{U}}_{new} \gets \nabla I(x,y)$ at scale s
\State ${\vec{U}}_{new} \gets \nabla c.\nabla I(x,y)+c(x,y,t)\Delta I$ 
\For{\texttt{<each pixel in image>}}
\If{$\parallel {\vec{U}}(x,y) \parallel < M/C $} \Comment{C is thresholding constant}
\State ${\vec{U}}(x,y)  = {\vec{U}}_{new}(x,y)  $
 \ElsIf{$abs(arctan({\vec{U}}(x,y), {\vec{U}}_{new}(x,y) )) < \pi/4$}
    \State ${\vec{U}}(x,y)  = {\vec{U}}(x,y) + {\vec{U}}_{new}(x,y)  $
  \Else
    \State ${\vec{U}}(x,y) = {\vec{U}}(x,y)$
    \EndIf  
\EndFor
\textbf{end \{if, for\}} 
\State The final ${\vec{U}} $ is the edge flow vector we are interested in
\State ${\vec{U}} \gets Binarize({\vec{U}})$
\State Enhance ${\vec{U}}$ by performing morphological operations
\For{\texttt{gradient magnitude image in scale $\textit{s}_n $}}
  \If{{$n\geq 2, use$ $strel -> 0px, disc$} 
  \Else		\Comment{structural element(strel)}
    \State $strel -> 1px, disc$} 
    \EndIf
    \State Apply  \textbf{Erosion} operation; \textbf{Close} to recover edges
\EndFor \textbf{end for} 
\State Obtain centroids from \textit{binary} edge-map
\State Obtain minBoundCircle and minBoundEllipse on centroids
\State \textbf{Calculate} Dice Coefficient \Comment{Iterate over scales}
\EndWhile \textbf{\\ end while} \\
The final boundary is given by minBoundCircle and\ or minBoundEllipse after GoF maximization. 
%\EndProcedure
\end{algorithmic}
\end{algorithm}
The algorithm searches the edges in finer scales and strengthen them with the edges recovered from higher scale. The homogeneous regions have vectors of zero length, so the detected edge segments grow in thickness (and often strength) as we move from lower to higher scales. Some edges do not exist in lower scale, but can still be significant. To decide on the same and reduce noise we put a boundary condition - and when the maximal edge strength (M) is greater than a heuristically predefined constant (C) -the edges are retained, or else they are discarded.The primary objective of applying a edge detection is to delineate the boundary of the imaged object and differentiate it from the background. But often in optoacoustics, the signal originate from the impurities or inhomogenities withing coupling medium. Further, noisy background is present in reconstructed images (lower boundary in Fig 1.a) due to limited view, and shortcomings of inversion methodologies. This noises are often strong enough to be detected by edge detection algorithm as true edges. Thus, we use an anisotropic diffusion process to further clean up the image, the diffusion process smoothens the image without suppressing the edges. Thereafter a non-linear morphological processing is done on the binary (diffused)edge mask. We take an sub-pixel sampling approach (0.5 px), rendering the operation is redundant beyond the second scale level. The morphological mask is differentially chosen at different scales. Initially, the image is eroded with a disc structuring element to remove noisy patches, but it also thins the edges. To recover the edges a closing operation is executed, with smaller structural element for erosion and a bigger element (2px) for dilation. 

\subsection{Parametric shape model fitting and Goodness of fit}
\label{sec:modelfit}

In section\ref{sse: edgeflow} we discussed about detection of edges, several authors have utilized the edge-flow vector for segmentation. In OAT images we see formation of smaller edge clusters and open contours. Thus, getting a ideal segmentation using edge linker seem to perform poorly. However, given the fact that our current problem which requires segmenting the image into only two classes - image and background, we follow a simple curve fitting approach. The proposed method first generates the centriods for edge clusters and then try to fit on a geometric pattern (deformable ellipse) iteratively through a set of parametric operations. A typical scenario in curve fitting is when the data is a best fitted, but some data points lies outside the curve, as we take an interpolated spline fitting criterion. In our approach, we modeled an inclusion criterion which enclosed all centroids and create forms a closed curve. Theoretically, it draws a convex hull with the centroids on a perimeter and approximate it to the nearest curve. In some datasets, we see presence of the outlier centroids which significantly biases the curve (due to presence of the inclusion criterion) leading to erroneous fitting. To avoid such complications, the values of the centroid positions are filtered for squeezing out outliers through a median filter, and then analyzed for generating the shape models which match the object boundary.The goodness of fit (GoF) is calculated using the average DM, which is a measure of contour overlap utilizing the area under the fitted curve (A), manually segmented region (M), and their intersection. DM always vary between 0 and 1, with DM $\geq 7$ being considered a good segmentation\cite{lynch2008segmentation}. The DM can be expressed as: 
\begin{equation}
DM = \frac{2\mid(A\cap M)\mid}{\mid A \mid + \mid M \mid}
\end{equation}
The GoF measure is used to iteratively improve the segmentation performance - the images are investigated at deeper scale spaces and the morphological diffusion step is decided based on the GoF value. We break away from the loop once a stable GoF is obtained and going to lower scale space negatively impact the calculated GoF values.The values for Rand Index (RI), which gives a measure of similarity and statistically used to quantify accuracy are also calculated and plotted in Figure 3. 

\section{Results and Discussion}
\label{sec:results}

For testing and standardization of the introduced image processing framework, a database of 30 datasets for 3 different anatomical regions in mice (10 for each region) was created. Each individual dataset represents signals (averaged 10 times) acquired at up to 6 different positions and at 8 different wavelengths (between 690-900 nm). Since manual segmentation of each dataset was needed for computing the performance of the algorithm vis-a-vis ground truth, the number of datasets was scaled down. Thereby, 4 datasets were considered from each anatomical regions, namely brain, liver and kidney/spleen, for computing the goodness of fit parameters. The proposed algorithm demonstrated good edge recovery performance. As shown in Fig 1, the algorithm weighted the edges that appear in multiple scales and reduced the spurious edges. The assumption made by \cite{ma2000edgeflow} is that if the vector direction change matches in multiple scales, then we can infer that it corresponds to real edges. Thus, the vector directions were checked in both finer and coarser scales, in a way that when there is a match, the designated edges were strengthened. The construction of the algorithm allows to detect the initial edge points from the finer scales, and reinforce then as we move to the larger scales. The finer scales are more immune to noise and the use of a non-negative constraint during the image reconstruction process prevents unnatural movement of the vector field (potentially due to absence of undesired negative values). Thus the edges detected in the finer scales are very significant to recover object boundaries, and are helpful in segmentation of OAT images. In Figure 2, we show the performance of multiscale segmentation (2c) along with the edge map recovered using Sobel operator (2b). The improvement in the edge detection performance by considering multiple scale (as over single scale in Sobel) is evident in the combined multiscale edge map ($\sigma = 1-3$). Further, a closer observation reveals that the morphological processing have successfully eliminated the spurious edges formed beyond the tissue boundary (Fig. 2d). Finally, in Figure 2(d-e) we show the calculated centroid clusters obtained from the morphologically processed binary edge map, and the ellipse fitting model applied to this centroids respectively.

Thereafter, we computed the goodness of fit using quantitative measures, viz. DM and RI, with ground truth (manual segmentation) as reference. In Table 1 we show the performances of the curve fitting using the DM (\ref{euclid}). Theoretically, DM values above 0.7 are considered to represent a good segmentation result, and using the proposed methodology DM values between 0.90 and 0.96 were achieved. Morphological processing and increasing scale-space depth was employed by tracking the corresponding changes in GoF value to improve noise performance when outlier centroids are present. Improvements in DM values was observed in most regions by adding anisotropic diffusion and morphological filtering with iterative GoF optimization, except for the kidney/spleen zone where the DM decreased after filtering, although DM values above 0.95 both with and without secondary filtering were achieved. From figure 3, it is be clearly inferred from both DM and RI that the improvements in the curve fit performance is observed in scale 2 followed by a marginal change in scale 3. So, for most practical purposes and reducing computational overhead, as scale depth of 2 can be chosen for OAT imaging experiments and image segmentation. 

% Below is an example of how to insert images. Delete the ``\vspace'' line,
% uncomment the preceding line ``\centerline...'' and replace ``imageX.ps''
% with a suitable PostScript file name.
% -------------------------------------------------------------------------

\begin{figure} [htb]
\begin{minipage}[b]{1\linewidth}
  \centering
 % \vspace{-1.5cm}
  \centerline{\includegraphics[width=6cm]{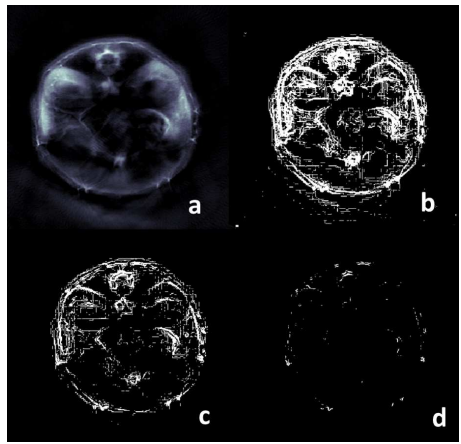}}
  %\vspace{0.8cm}
%  \vspace{1.5cm}
%  \centerline{(a)Edge maps from different operators}\medskip
\end{minipage}
\caption{ Reconstructed optoacoustic image using non-negative constrained model based inversion (a), and (b-d) edges detected at multiple scales $s =1,2,3$ respectively.}
\label{Fig1}
\end{figure}

\begin{figure} [htb]
\begin{minipage}[b]{1\linewidth}
  \centering
  \vspace{-0.5cm}
  \centerline{\includegraphics[width=7cm]{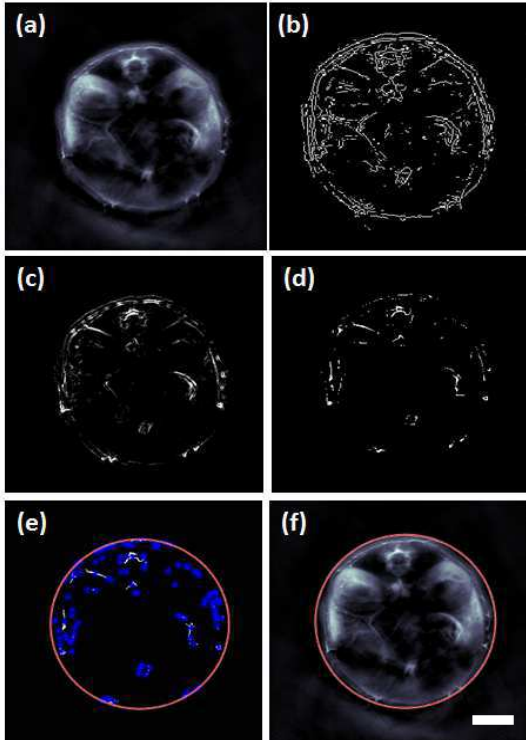}}
  \vspace{-0.2cm}
%  \vspace{1.5cm}
%  \centerline{(a)Edge maps from different operators}\medskip
\end{minipage}
\caption{ Reconstructed Optoacoustic image with non-negative constrain(a), edge-maps generated by Sobel operator (b), Multiscale (Edgeflow) map  (c)and (proposed) Morphological processed multiscale edge map (d) are shown.  In (e) computed centroids from edge clusters are shown in blue,  and (f) illustrates the curve fitting method applied to the centroids (fitted curve marked in red). [Scale = 3mm]}
\label{fig:res}
\end{figure}
\vspace{-0.1cm}
\centerline{\textbf{Table 1.} Efficacy of Curve Fitting}
\scalebox{1.0}{
\begin{tabular*}{0.45\textwidth}{lllr}
\hline
\multicolumn{3}{r}{Dice Coefficient Metrics} \\
\cline{2-3}
Regions        &  Non-optim   &  GoF-Optim  & Rand Index \\
\hline
Brain           & 0.9387 & 0.9440 &0.9771  \\
Liver           & 0.9093 & 0.9447 &0.9396 \\
Kidney/Spleen   & 0.9606 & 0.9571 &0.9672\\

\hline
\end{tabular*}}

\section{Conclusions}
\label{sec:illust}

In this article, a method of delineating the boundary of optoacoustic small animals images using a multiscale edge detection algorithm in combination with geometrical curve fitting has been presented. It is noteworthy that the method employed is self-deterministic and requires minimal human intervention. The optoacoustic signal/image datasets can be very large given the 5D nature of this modality \cite{7106682}, automating the image formation and analysis workflows is a very challenging and important problem. Thus, the algorithms and the workflows demonstrated herein are expected to be helpful in automating optoacoustic image segmentation, with important significance towards enabling quantitative imaging applications.
\begin{figure}[htb]
\label{OAT_prinzip}
\begin{minipage}[b]{1.0\linewidth}
  \centering
  \vspace{-0.5cm}
  \centerline{\includegraphics[height=5cm, width=9cm,]{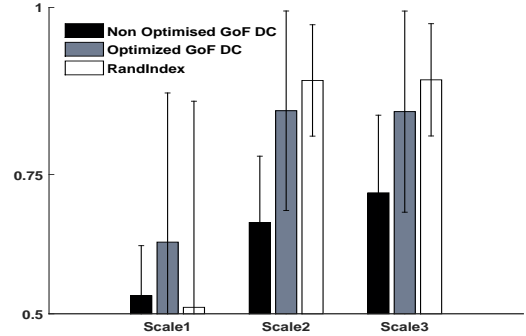}}
  \vspace{-0.5cm}
  \end{minipage}
\caption{GoF trace and the standard deviation for the torso(kidney/spleen region) scans of mouse. (Total dataset size = 12 (z-slices in whole body small animal tomographic scanner), acquired from 2 different mice) \textit{in-vivo}}
\end{figure}

\section{Acknowledgments}
\label{sec:ack}

S.M. acknowledges support (PhD Scholarship Award: A/11/75907) from the German Academic Exchange Service (DAAD). D.R. acknowledges support from the European Research Council under grant agreement ERC-2010-StG-260991.


\bibliographystyle{IEEEbib}
\bibliography{refs}

\begin{thebibliography}{10}

\bibitem{razansky2011volumetric}
D~Razansky, A~Buehler, and V~Ntziachristos,
\newblock ``Volumetric real-time multispectral optoacoustic tomography of
  biomarkers,''
\newblock {\em Nature protocols}, vol. 6, no. 8, pp. 1121--1129, 2011.

\bibitem{sumengen2005multi}
B~Sumengen and BS~Manjunath,
\newblock ``Multi-scale edge detection and image segmentation,''
\newblock in {\em European signal processing conference (EUSIPCO)}. Citeseer,
  2005.

\bibitem{canny1986computational}
J~Canny,
\newblock ``A computational approach to edge detection,''
\newblock {\em Pattern Analysis and Machine Intelligence, IEEE Transactions
  on}, , no. 6, pp. 679--698, 1986.

\bibitem{perona1990scale}
P~Perona and J~Malik,
\newblock ``Scale-space and edge detection using anisotropic diffusion,''
\newblock {\em Pattern Analysis and Machine Intelligence, IEEE Transactions
  on}, vol. 12, no. 7, pp. 629--639, 1990.

\bibitem{Mandal2014128}
S~Mandal, E~Nasonova, XL~Deán-Ben, and D~Razansky,
\newblock ``Optimal self-calibration of tomographic reconstruction parameters
  in whole-body small animal optoacoustic imaging,''
\newblock {\em Photoacoustics}, vol. 2, no. 3, pp. 128 -- 136, 2014.

\bibitem{tabb1997multiscale}
M~Tabb and N~Ahuja,
\newblock ``Multiscale image segmentation by integrated edge and region
  detection,''
\newblock {\em Image Processing, IEEE Transactions on}, vol. 6, no. 5, pp.
  642--655, 1997.

\bibitem{ma2000edgeflow}
WY~Ma and BS~Manjunath,
\newblock ``Edgeflow: a technique for boundary detection and image
  segmentation,''
\newblock {\em Image Processing, IEEE Transactions on}, vol. 9, no. 8, pp.
  1375--1388, 2000.

\bibitem{6748955}
A.~Taruttis, A.~Rosenthal, M.~Kacprowicz, N.C. Burton, and V.~Ntziachristos,
\newblock ``Multiscale multispectral optoacoustic tomography by a stationary
  wavelet transform prior to unmixing,''
\newblock {\em Medical Imaging, IEEE Transactions on}, vol. 33, no. 5, pp.
  1194--1202, May 2014.

\bibitem{10.1109/TPAMI.1987.4767980}
F~Bergholm,
\newblock ``Edge focusing,''
\newblock {\em IEEE Transactions on Pattern Analysis and Machine Intelligence},
  vol. 9, no. 6, pp. 726--741, 1987.

\bibitem{lynch2008segmentation}
M~Lynch, O~Ghita, and PF~Whelan,
\newblock ``Segmentation of the left ventricle of the heart in 3-d+ t mri data
  using an optimized nonrigid temporal model,''
\newblock {\em Medical Imaging, IEEE Transactions on}, vol. 27, no. 2, pp.
  195--203, 2008.

\bibitem{7106682}
S.~Mandal, X.L. Dean-Ben, N.C. Burton, and D.~Razansky,
\newblock ``Extending biological imaging to the fifth dimension: Evolution of
  volumetric small animal multispectral optoacoustic tomography.,''
\newblock {\em Pulse, IEEE}, vol. 6, no. 3, pp. 47--53, May 2015.

\end{thebibliography}

\end{document}